\documentclass[fleqn, usenatbib]{mnras}

\usepackage{newtxtext,newtxmath}

\usepackage[T1]{fontenc}
\usepackage{ae,aecompl}

\usepackage{graphicx}	
\usepackage{amsmath}	
\usepackage{amssymb}	
\usepackage{cases}

\def\aj{AJ}
\def\araa{ARA\&A}
\def\apj{ApJ}
\def\apjl{ApJ}
\def\apjs{ApJS}
\def\aap{A\&A}
\def\mnras{MNRAS}
\def\nat{Nature}

\title[Bar signatures on stellar kinematics]{Signatures of the Galactic bar on stellar kinematics unveiled by APOGEE}

\author[Palicio et al.]{
Pedro A. Palicio,$^{1,2}$\thanks{E-mail: pedroap @ iac.es}
Inma Martinez-Valpuesta,$^{1,2}$
Carlos Allende Prieto,$^{1,2}$ \newauthor
Claudio Dalla Vecchia,$^{1,2}$
Olga Zamora,$^{1,2}$
Gail Zasowski,$^{3,4}$
J. G. Fernandez-\newauthor Trincado,$^{5,6}$
Karen L. Masters,$^{7,8}$
D. A. Garc\'ia-Hern\'andez,$^{1,2}$
Alexandre\newauthor Roman-Lopes,$^{9}$\\
\\
$^{1}$Instituto de Astrof\'isica de Canarias, E-38205 La Laguna, Tenerife, Spain\\
$^{2}$Universidad de La Laguna, Dpto. Astrof\'isica, E-38206 La Laguna, Tenerife, Spain\\
$^{3}$Space Telescope Science Institute, 3700 San Martin Drive, Baltimore, MD 21218, USA\\
$^{4}$Department of Physics and Astronomy, University of Utah, Salt Lake City, UT 84105, USA\\
$^{5}$Departamento de Astronom\'ia, Universidad de Concepci\'on, Casilla 160-C, Concepci\'on, Chile\\
$^{6}$Institut Utinam, CNRS UMR6213, Univ. Bourgogne Franche-Comt\'e, OSU THETA , Observatoire de Besan\c{c}on, BP 1615,\\ 25010 Besan\c{c}on Cedex, France\\
$^{7}$Institute of Cosmology and Gravitation, University of Portsmouth, Dennis Sciama Building, Portsmouth, PO1 3FX, UK\\
$^{8}$Haverford College, Department of Physics and Astronomy, 370 Lancaster Avenue, Haverford, Pennsylvania 19041, USA\\
$^{9}$Departamento de F\'isica, Facultad de Ciencias, Universidad de La Serena, Cisternas 1200, La Serena, Chile
}

\date{Accepted XXX. Received YYY; in original form ZZZ}

\pubyear{2018}

\begin{document}
\label{firstpage}
\pagerange{\pageref{firstpage}--\pageref{lastpage}}
\maketitle

\begin{abstract}
Bars are common galactic structures in the local universe that play an important role in the secular evolution of galaxies, including the Milky Way. In particular, the velocity distribution of individual stars in our galaxy is useful to shed light on stellar dynamics, and provides information complementary to that inferred from the integrated light of external galaxies. However, since a wide variety of models reproduce the distribution of velocity and the velocity dispersion observed in the Milky Way, we look for signatures of the bar on higher-order moments of the line-of-sight velocity ($V_\textnormal{los}$) distribution. We make use of two different numerical simulations --one that has developed a bar and one that remains nearly axisymmetric-- to compare them with observations in the latest APOGEE data release (SDSS DR14). This comparison reveals three interesting structures that support the notion that the Milky Way is a barred galaxy. A high skewness region found at positive longitudes constrains the orientation angle of the bar, and is incompatible with the orientation of the bar at $\ell=0^\circ$ proposed in previous studies. We also analyse the $V_\textnormal{los}$ distributions in three regions, and introduce the Hellinger distance to quantify the differences among them. Our results show a strong non-Gaussian distribution both in the data and in the barred model, confirming the qualitative conclusions drawn from the velocity maps. In contrast to earlier work, we conclude it is possible to infer the presence of the bar from the kurtosis distribution.
\end{abstract}



\section{Introduction}
\par Since bars are present in about two thirds of the local disc galaxies \citep{Eskridge2000, MarinovaJogee2007, Sellwood2014}, the study of such structures is key to understand galaxy evolution. Among others, radial migration along bars has been suggested as a mechanism to grow central bulges, and bar fraction correlates with the cessation of star formation in galaxies in the local Universe \citep{Masters_et_al2010, Masters_et_al2012, Wang_et_al2012, Cheung_et_al2013, Gavazzi_et_al2015, Spinoso_et_al2017}. Fortunately, the Milky Way offers an extraordinary opportunity to shed light on these processes, given that it is thought to have a bar, as initially suggested by \citet{deVaucoul1964} to explain the non-circular motions observed in \textsc{Hi}, and subsequently confirmed by the near-IR images of \citet{Matsumoto_et_al1982} \citep{BlitzSpergel1991}, the \textit{COBE}/DIRBE maps \citep{Weiland1994, Dwek1995, BinneyGerhardSpergel1997}, star counts \citep{Nakada_et_al1991, Whitelock_Catchpole1992, Weinberg1992, Stanek_et_al1994, StanekUdalski1997, Hammersley2000, Benjamin2005, Robin_et_al2012, WeggGerhardPortail2015}, the distribution of globular clusters \citep{Blitz1993}, and the rate of microlensing events in the Galactic bulge \citep[][but see also \citealt{KiragaPacz1994, HanGould1995, Alcock_et_al1997, StanekUdalski1997}]{Paczynski1994, Evans1994, ZhaoSpergelRich1995}.
\par The search for the Galactic bar has also been attempted using stellar kinematics. \citet{ZhaoSpergelRich1994} proved that the vertex deviation of 62 K-giant stars in the Baade's Window is inconsistent with an axisymmetric bulge potential. \citet{Rattenbury2007b} found a longitudinal asymmetry in the proper motion dispersions of 45 OGLE-II fields that may be due to the bar (see their figures 3 and 4). \citet{Dehnen2000} identified a bimodality in the velocity distribution of 14,000 \textit{Hipparcos} stars \citep{Perryman1997, ESAHipparcosTycho} caused by the outer Lindblad resonance (OLR). In a similar study, \citet{GarnerFlynn2010} reproduced the Hercules stream both with a standard Galactic bar and a ``long bar'' \citep{Benjamin2005, LopezCorr_et_al2007}, although its position in the velocity plane shows a weak dependence with the orientation angle. On the contrary, \citet{PerezVillegas_et_al17} proposed an alternative explanation in which the Hercules stream is mainly composed by stars orbiting the stable Lagrangian points, which energy is high enough to visit the solar neighbourhood. However, the OLR reproduces better the bimodality observed in the distributions of the line-of-sight velocities at the galactic longitude $\ell=270^\circ$ and latitude $b=0^\circ$ \citep{Hunt_et_al2017}.
\par Apart from the Hercules stream, other moving groups of stars such as the Hyades, Sirius, and the Pleiades show kinematic features that can be explained with the gravitational effect of a bar potential \citep{Kalnajs1991, Minchev_et_al2010}. In particular, \citet{Minchev_et_al2010} reproduce either the Sirius or the Coma Berenices moving groups depending on the orientation angle and the time elapsed after the bar formation. These streaming motions may not be the unique imprint of the bar on the velocity distribution, since, as \citet{Molloy_et_al15} suggested, the high velocity peaks detected by \citet{NideveretalHVP2012} and confirmed by \citet{Zasowski_et_al16} may be a consequence of the 2:1 orbits that support the bar (though other alternatives have been proposed: \citet{LiShenRichHVP,DebattistaNessEarpHVP2015,Aumer_Schonrich15}). In a recent work, \citet{BobylevBajkova17} reported changes in the orbits of ten globular clusters due to the inclusion of the bar potential, although this result could be more affected by the uncertainties in the input.
\par Regarding the Oort constants, several authors find values for \textit{C} and \textit{K} incompatible with axisymmetric potentials in which \textit{C}=\textit{K}=0 km s$^{-1}$ kpc$^{-1}$ \citep{Comeron1994, Torra_et_al2000, OllingDehnen2003, Minchev_et_al2007, Bovy_GaiaDR1_2017}. In particular, \citet{Bovy_GaiaDR1_2017} used the TGAS catalogue, included in the first \textit{Gaia} Data Release \citep{TGAS25, GaiaDR1, GaiaColab2016, GaiaCollaborationDR1} to provide the most accurate estimation of \textit{C} and \textit{K} to date (\textit{C}= $-3.2\pm0.4$ km s$^{-1}$ kpc$^{-1}$ and \textit{K}= $-3.3\pm0.6$ km s$^{-1}$ kpc$^{-1}$) and confirm their non-zero values. On the other hand, \citet{Minchev_et_al2007} not only explained the differences between the cold disc stars (\textit{C}$\approx0$ km s$^{-1}$ kpc$^{-1}$) and the hot disc component (a significantly negative value of \textit{C}) reported by \citet{OllingDehnen2003}, but also observed deviations in the Oort constants \textit{A} and \textit{B} due to the bar. 
\par Despite these previous studies confirm the effects of the bar on the Galactic kinematics, the distributions of the mean line-of-sight velocity and dispersion in the Milky Way can be reproduced with several models, which hinders the determination of some parameters key to understanding the characterization and evolution of the Milky Way, such as the bar length and pattern speed, $\Omega_p$. In order to break this degeneracy, \citet{Zasowski_et_al16} studied the higher-order kinematic moments of the APOGEE DR12 stars \citep{APOGEEDR12paper} and compared them with the simulations of \citet{MartValGerh2011} and \citet{Shenetal2010}. Their results show structures in the skewness of the line-of-sight velocity, $V_\textnormal{los}$, especially for the metal-rich stars, but a nearly flat pattern for the kurtosis.
\par In this work, we aim to explore the imprints of the bar on the higher-order moments of $V_\textnormal{los}$ by using the in-plane projection (galactic longitude vs. distance projected on the Galactic plane; that will be introduced in a companion paper \citet{MyPaper}, in preparation) and the distributions of $V_\textnormal{los}$. We assume the solar abundances provided by \citet{Asplund_et_al2005} as a reference for the metallicity ($Z_\odot=0.0122$). We investigate the maps of the simulations and observations to find the bar structures and compare them in qualitative terms. This comparison is complemented by a quantitative analysis of the velocity distributions provided by the Hellinger distances.
\par This paper is organised as follows: the observational data is described in section \ref{Real_data_section} while the simulations are explained in section \ref{Sim_data_section}. The maps and distributions of the line-of-sight velocities are introduced in section \ref{Section_Results} and discussed in \ref{Discussion_Section}. Finally, the conclusions are included in section \ref{Summary_Section}.
%
%
%
%
\section{Observational Data and distance estimation}
\label{Real_data_section}
\par The Apache Point Observatory Galactic Evolution Experiment \citep[APOGEE;][]{APOGEEpaper, APOGEE2paper} is a homogeneous spectroscopic survey in the near-infrared \textit{H} band (1.51-1.70 $\mu$m) included as part of the SDSS-III project \citep{SDSSIIIpaper}, and continuing in SDSS-IV \citep{SDSSIVpaper}. It provides high resolution spectra ($R\sim 22500$) for stars in all the Milky Way components, with special emphasis on dust-obscured regions such as the disc and bulge, allowing a detailed study of the chemistry and kinematics in the inner Galactic regions.
\par Since the observations are carried out with the Sloan 2.5-m Telescope \citep{Gunn_et_al_25mTelescope} at Apache Point Observatory (New Mexico), there is a significant unexplored region at negative longitudes in the APOGEE survey that restricts the study of the inner Galactic kinematics to the receding part of the Milky Way. This limitation will be overcome with the observations from the du Pont 2.5-m Telescope at Las Campanas Observatory (Chile), as part of the oncoming extension of the APOGEE survey \citep[APOGEE-2; ][but see also \citealt{APOGEE2S_target} for the planned field and target selection]{APOGEE2paper}, which first dual-hemisphere data release is expected by July 2018 \citep{SDSSIVpaper}.
\par The APOGEE DR14 \citep{preprintAPOGEEDR14} is the most recent of the APOGEE data sets, and provides stellar parameters, chemical abundances \citep{ASPCAPpaper} and line-of-sight velocities\footnote{To avoid confusion, we reserve the term ``radial velocity'' for the approaching/receding motion with repect to the Galactic center, while the line-of-sight velocity is the projection of the star motion along the visual line (i. e. $\ell$ and $b$ remain fixed). } for more than 250,000 sources, doubling the number of stars with respect to its antecessor DR13 \citep{APOGEEDR13paper}. In addition to that, both data sets include photometry (\textit{J}, \textit{H} and \textit{Ks} bands) and the sky coordinates ($\ell$, $b$) of the 2MASS sources \citep{TWOMASSpaper}.
\par We make use of the distances calculated with the Bayesian method developed by \citet{BPG_distances} and \citet[][in preparation]{Queiroz_et_al17}, in which an input of observed photometric and spectroscopic parameters is compared to stellar models to return the posterior probability of the distance distribution. The median value of this distribution is adopted as estimate of the distance, with typical uncertainties for the giant stars of about 20\%. We discard a crossmatch with the first Gaia data release because the TGAS solution does not measure parallaxes accurately enough for stars beyond $\sim 0.3$ kpc.

\par In terms of kinematics, APOGEE provides lines-of-sight velocities with precision under 0.1 km s$^{-1}$ \citep{Nidever_et_al2015VLOS}. We correct heliocentric velocities by subtracting the solar motion according to the following equation:
\begin{equation}
 \label{Solar_Motion_correction}
 V_\textnormal{los,gal} = V_\textnormal{los,helio} + \left( V_{\odot,\textnormal{y}}\textnormal{cos}\ell - V_{\odot,\textnormal{x}}\textnormal{sin}\ell\right) \textnormal{cos}b + V_{\odot,\textnormal{z}}\textnormal{sin} b
 \end{equation}
where $\vec{V}_\odot$=(-241.92, 11.1, 7.25) km s$^{-1}$ \citep{ReidBrunthaler04, SchBinDeh2010}.
\par It is worthwhile noting that this correction does not depend on the heliocentric distances, but only on the 2MASS coordinates ($\ell$, $b$), whose errors are negligible for our purposes.
\par We select APOGEE sources within $-6^\circ\leq\ell<42^\circ$ to include the near bar arm, and impose the additional cutoff $|Z|>1$ kpc in the vertical direction to exclude the stars outside the disc. We cannot consider a symmetric range in $\ell$ because the latitude of Apache Point Observatory makes it impossible to observe the inner Galaxy at negative longitudes. All the stars with bad flags in their stellar parameters, or no distance estimation, are discarded. Foreground sources are excluded by imposing $d\cos{b}\geq 3$ kpc, while different upper limits in distance are considered. These restrictions lead to a final sample size of $\sim 11000$ sources with $d\cos{b}\in [3, 12)$ kpc. 
%
%
\section{Simulation data description}
\label{Sim_data_section}
\par We make use of the numerical simulations introduced in \cite{MyPaper} and compare them with observations. The simulations account for star formation and consider an exponential disc with a Toomre Q parameter of 1.5 embedded in a dark matter halo as initial conditions. The two models considered differ only in the fraction of baryonic matter: the disc of each galaxy contains 30\% and 50\% of the total mass within 7 kpc from the center, respectively. After a total simulation time of 2.52 Gyrs, the galaxy with larger fraction of baryonic matter has developed a $\sim4.5$ kpc (half-length) bar \citep{Benjamin2005, CLavers_et_al2007, MartValGerh2011, WeggGerhardPortail2015} and a pattern speed $\Omega_p\approx30$ km s$^{-1}$ kpc$^{-1}$, similar to that measured by \citet{PortailWeggGerhardMV15}. The other simulated galaxy does not develop a bar, and remains almost axisymmetric during the total simulation time of 4.48 Gyrs. In both simulations, the solar position is defined by the orientation angle $\phi_\textnormal{bar}=25^\circ$ \citep{StanekUdalski1997, Freudenreich1998, LopezCorr2005, Rattenbury2007, Shenetal2010, WeggGerhard2013, Caoetal2013, Natafetal2015}, with the positive $y$-semiaxis pointing in the $\ell=0^\circ$ direction. In this frame, the Sun is located at ($X$, $Y$, $Z$)= ($0$, -$R_0$, $Z_0$) with $R_0=8.0$ kpc and $Z_0=0.025$ kpc. Since the velocities are expressed in a galactocentric reference system, no solar motion correction is applied.
\par For each model, stellar velocities are rescaled using the factor $\lambda$ that minimizes the right-hand side of the following equation:
\begin{equation}
\label{eq_minimize}
 S = \sum_{bins} \left( V_\textnormal{los}^\textnormal{obs}-\lambda V_\textnormal{los}^\textnormal{sim}\right)^2+ \sum_{bins}\left(\sigma_\textnormal{los}^\textnormal{obs}-\lambda \sigma_\textnormal{los}^\textnormal{sim}\right)^2
\end{equation}
\par Using this method, we account for the line-of-sight velocity dispersion, which is much smaller in the axisymmetric model, and avoid a fit based only on the $\langle V_\textnormal{los} \rangle$ map, that would lead to an unrealistically low $\sigma_\textnormal{los}$. Furthermore, as the maps for the non-axisymmetric model contain more structure due to the bar, the scale factor should not be a bare average.
\par Our work is inspired by \citet{Abbott_et_al17} who also took into account the dispersion for the fitting of the disc mass to match the results of the BRAVA survey \citep{Rich_et_al2007, Kunder_et_al_BRAVAII2012}. Other statistical parameters, such as the skewness and kurtosis are not included in equation \ref{eq_minimize} because they are dimensionless.
\section{Results}
\label{Section_Results}
\subsection{Velocity maps for different stellar metallicities}
%
%
%
\begin{figure*}
\begin{center}
\includegraphics[scale = .5]{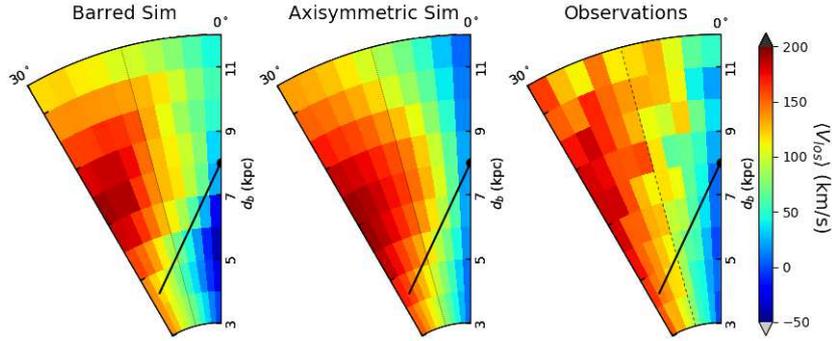}
\caption{Maps of the averaged line-of-sight velocities for the barred model (left), axisymmetric model (middle) and the observations (right) in the region $0^\circ\leq \ell <30^\circ$ and $3\leq d\textnormal{cos}b<12$ kpc.  The solid black line represents the bar major axis and the Galactic Center is denoted by the black spot at $\ell=0^\circ$, $d_b=8.0$ kpc.}
\label{Pict_BAO}
\end{center}
\end{figure*}
\par We use the maps introduced by \citet{MyPaper} to show face-on galactic projections since the uncertainties in distances change only the positions along one axis, keeping the line-of-sight fixed, in contrast to the usual projection in the $x-y$ plane. In addition, the increasing bin size with distance compensates the lack of observed sources at far distances, making the statistics more robust.
\par As can be seen in Fig. \ref{Pict_BAO}, the numerical models and the observations show similarities in their $\langle V_\textnormal{los}\rangle$ maps even at large heliocentric distances, where the APOGEE detection is biased by more metal-poor stars being brighter and easier to observe \citep{Hayden_et_al2015}. Indeed, stars with low [Fe/H] tend to be kinematically hotter \citep{Babusiaux_et_al10, ConjoinedRC_DePropris_et_al2011, NessFreeAth2012, NessFreeAthaWylie_ARGOSIII2013, NessFreeAthaWylie_ARGOSIV2013, GaiaESO_RojasA_et_al14, DiMatteo_et_al_2015, NessZasowJohn2016, Kunder_et_al2016, DiMatteo2016, Babusiaux2016, Williams_et_al2016, Zoccali2017, PortailWeggGerhardNess2017}, and the observed trends in $\langle V_\textnormal{los} \rangle$ may be a consequence of this.
\par
\begin{figure*}
\begin{center}
\includegraphics[scale = .5]{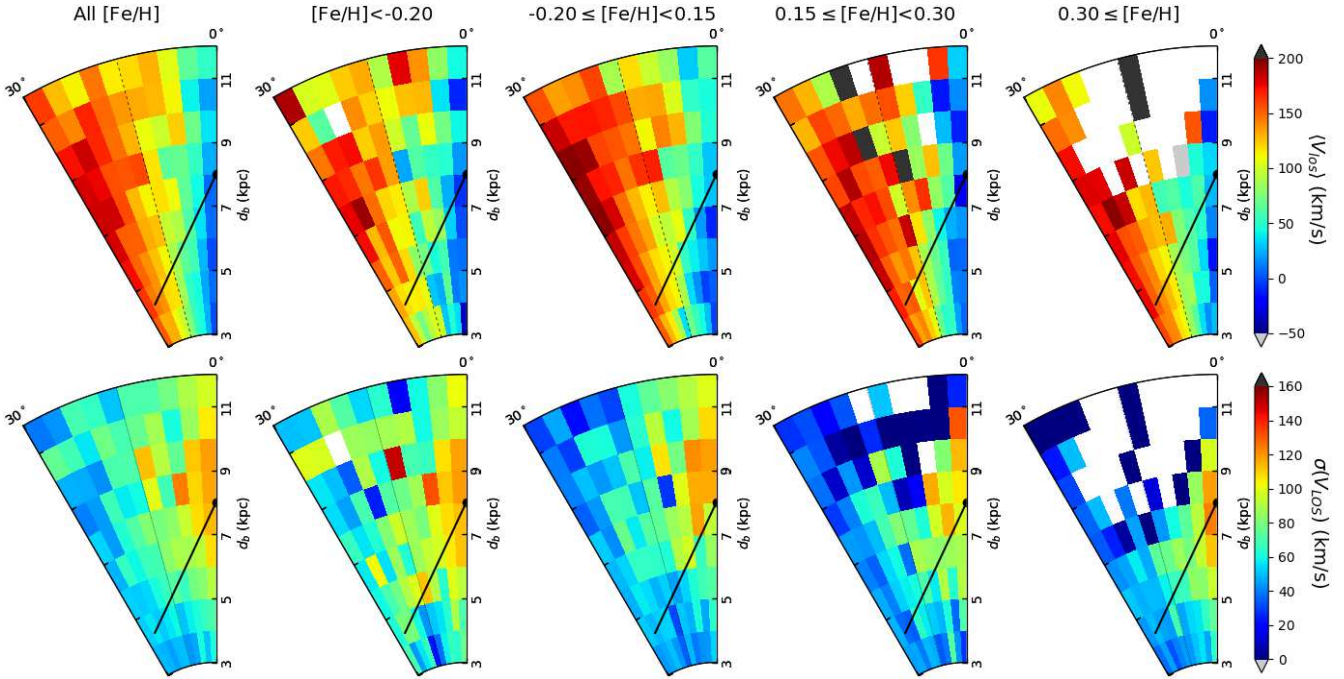}
\caption{Maps of the averaged line-of-sight velocities (upper row) and velocity dispersion (lower row) for the different metallicity ranges, each of them given by the quartiles of the [Fe/H] distribution for $0^\circ\leq \ell <30^\circ$ and $3\leq d\textnormal{cos}b<12$ kpc. The numbers of sources in each metallicity bin are (from left to right): 8314, 2207, 2180, 1654 and 2273 stars.}
\label{Pict_AllFeH}
\end{center}
\end{figure*}

 In order to quantify this bias, we grouped stars in four metallicity bins according to the quartiles of the [Fe/H] distribution in the region $0^\circ\leq\ell<30^\circ$ and $3\leq d\textnormal{cos}b<12$ kpc. As shown in Fig. \ref{Pict_AllFeH}, the $\langle V_\textnormal{los}\rangle$ and $\sigma_{los}$ maps are similar for $d\lesssim9$ kpc, while at further distance metal-rich stars become less abundant. The number of stars per bin decreases with distance for all metallicities, and ranges from $30-50$ in the closest regions to less than $15$ beyond $8$ kpc. The regions near the Galactic Centre are the most populated areas, with more than $40$ stars per bin for all the [Fe/H] values. In terms of kinematics, the largest discrepancies are observed at $15^\circ\lesssim\ell\lesssim30^\circ$ and $d>10$ kpc, where metal-poor stars ([Fe/H]$<-0.20$ dex) have lower $\langle V_{los} \rangle$. Comparing the maps of the dispersions, it is possible to detect a weak decrease in $\sigma_{los}$ with [Fe/H]. However, this trend is not strong enough to distort the global pattern. In order to reduce the bias, we restrict the study to the range $3\leq d_b<9$ kpc (hereafter $d_b \equiv d \textnormal{cos}b$ is the projection of the distance on the galactic plane).
%
%
%
%
\subsection{Maps of the mean values and dispersion}
\par
\begin{figure*}
\begin{center}
\includegraphics[scale = .62]{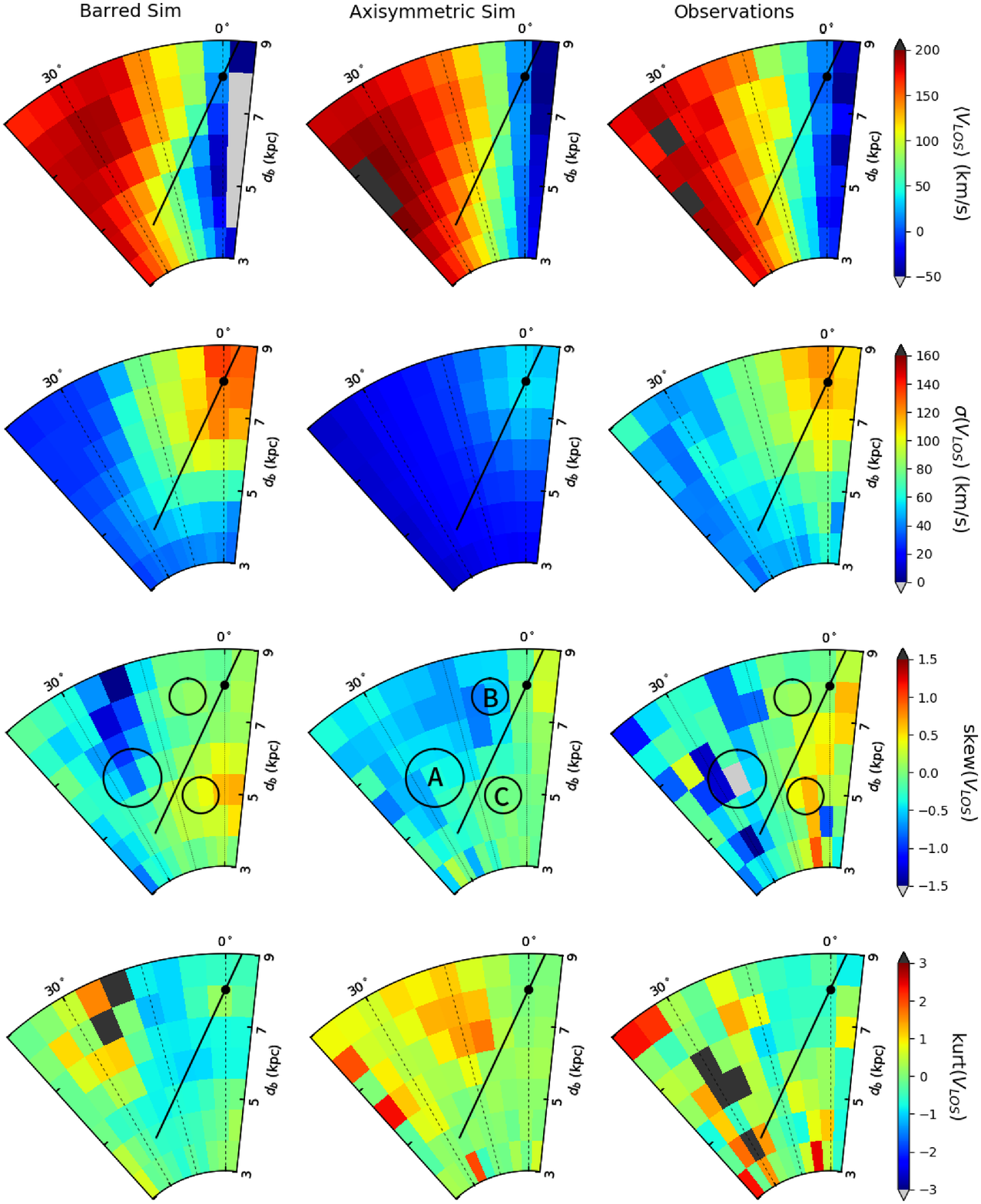}
\caption{Solid maps of the statistic parameters of $V_\textnormal{los}$ in the range $-6^\circ \leq \ell <42^\circ$, $d_b\in [3, 9)$ kpc and $|Z|<1$ kpc. From top to bottom: mean value  $\langle V_\textnormal{los} \rangle$, standard deviation  $\sigma_\textnormal{los}$, skewness (skew($V_\textnormal{los}$)) and kurtosis (kurt($V_\textnormal{los}$)) of the line-of-sight velocity. From left to right: barred model, axisymmetric model and observations. The Galactic center is placed at $\ell=0^\circ$, $d_b=R_0=8.0$ kpc (solid black circle) and the bar is plotted with an orientation angle of $25^\circ$ (black line). The open circles enclose Regions A, B and C discussed in the text.}
\label{Pict_Solid_full_ld}
\end{center}
\end{figure*}

\begin{figure*}
\begin{center}
\includegraphics[scale = .62]{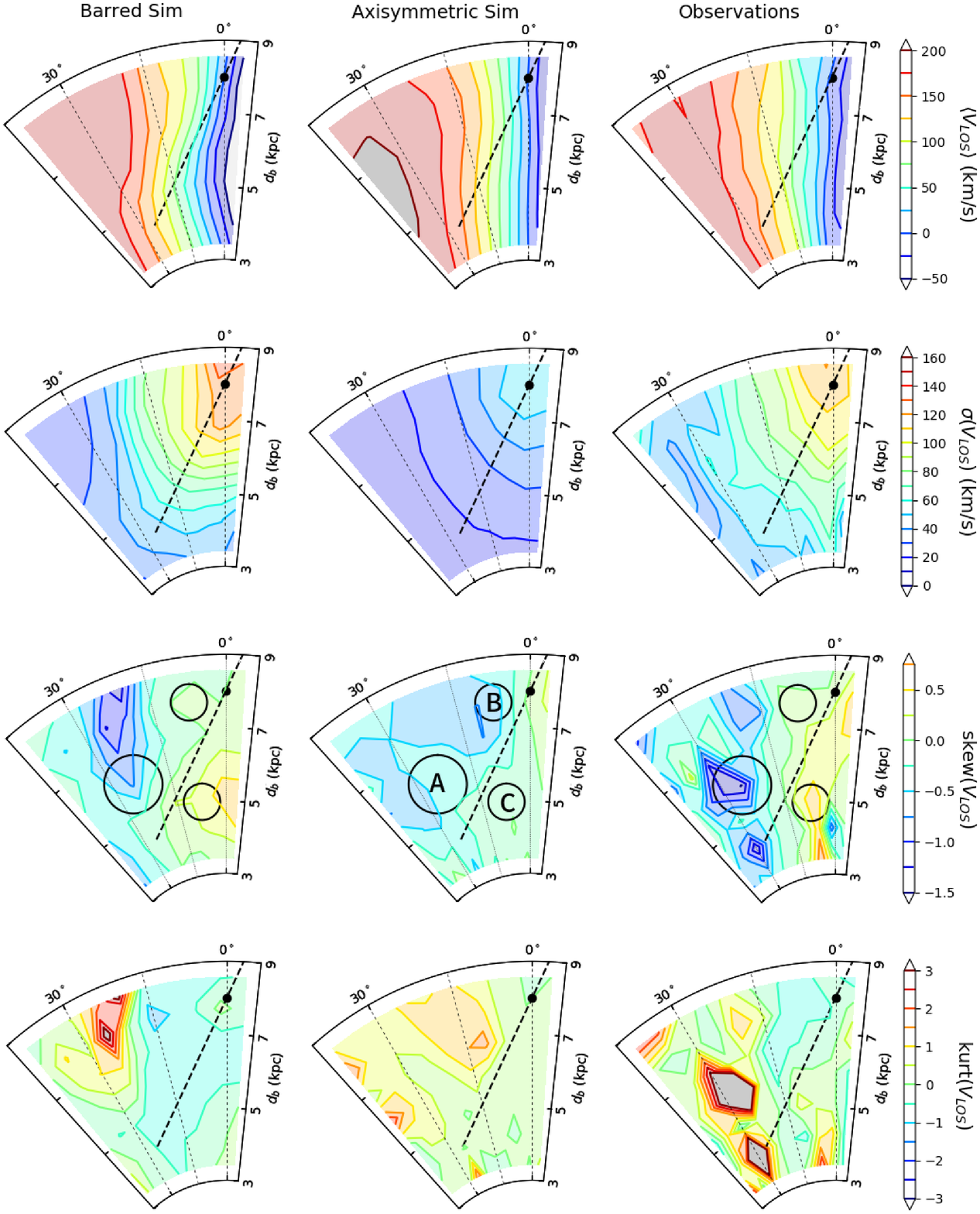}
\caption{Contour levels for the maps of Fig. \ref{Pict_Solid_full_ld} using the same convention. From top to bottom: mean value  $\langle V_\textnormal{los} \rangle$, standard deviation  $\sigma_\textnormal{los}$, skewness (skew($V_\textnormal{los}$)) and kurtosis (kurt($V_\textnormal{los}$)) of the line-of-sight velocity. From left to right: barred model, axisymmetric model and observations. The Galactic center is placed at $\ell=0^\circ$, $d_b=R_0=8.0$ kpc (solid black circle) and the bar is plotted with an orientation angle of $25^\circ$ (black line). The open circles enclose Regions A, B and C discussed in the text.}
\label{Pict_Contourf_full_ld}
\end{center}
\end{figure*}

 The maps of the mean value of $V_\textnormal{los}$ (first row in Figs. \ref{Pict_Solid_full_ld} and \ref{Pict_Contourf_full_ld}) show the characteristic velocity pattern expected for a clockwise-rotating system, with a very weak distortion in the contour lines for the barred case. Whereas the axisymmetric model shows less structure and a decreasing $\sigma_\textnormal{los}$ with galactocentric distance, a kinematically hotter elliptical enhancement is observed in the barred model. Compared to the numerical models, the Milky Way data show a more irregular enhancement aligned with the $\ell\approx0^\circ$ direction, with slightly lower $\sigma_\textnormal{los}$ values. 
\par Bayesian analysis, with the assumption of a Gaussian distribution of velocities in each bin, would favour the axisymmetric model due to the larger posterior probability given by its lower dispersion. However, as we show in the next section, this assumption does not hold for the barred model.
\par In order to discard the low source number as an explanation for the observed features, we examine the density of stars for the simulations and the observations. The simulations contain at least 230 particles per bin with a maximum density of $\sim 95 000$ particles in the regions close to the galactic center.  The observations, however, show a more heterogeneous distribution with highly populated bins ($>100$ stars per bin) along the bar and along the $\ell\approx 30^\circ$ direction up to $\sim 7$ kpc. With the exception of few regions at $\ell \gtrsim 30^\circ$ or $\ell \lesssim 0^\circ$, any of the bins contains at least $30$ stars. The minimum number of stars ($18$) is found in the most distant bin from the Sun ($\ell = 39.6$, $d_b \approx 8.6$ kpc). 
%
%
%
%
%
%
%
\subsection{Maps of skewness and kurtosis}
\label{subsec_skew_kurt}
\par We extend our study to the skewness and kurtosis of the velocity distribution to determine with more accuracy which model matches the Milky Way data best. The skewness involves the third moment of the distribution and provides information about the asymmetry of the distribution, while the kurtosis requires the fourth order moment and quantifies the flatness of the distribution. We calculate these parameters using
\begin{eqnarray}
 \label{eq_skew}
 \textnormal{skew}(x)&=&\frac{\langle \left(x-\langle x \rangle \right)^3\rangle}{\sigma^3}\\
 \label{eq_kurt}
 \textnormal{kurt}(x)&=&\frac{\langle\left(x-\langle x \rangle \right)^4\rangle}{\sigma^4}-3
\end{eqnarray}
where the rightmost term of eq. \ref{eq_kurt} accounts for the kurtosis of the Gaussian distribution\footnote{Sometimes this quantity is referred as \textit{excess of kurtosis}.}. It is worth mentioning that both values do not depend on the rescaling factor $\lambda$, and are dimensionless.
\par As Figs. \ref{Pict_Solid_full_ld} and \ref{Pict_Contourf_full_ld} illustrate, there are significant differences between the skewness distributions of the models. For the barred simulation we observe a low-skewness ($\lesssim-0.75$) banana-shaped region at $15^\circ<\ell<30^\circ$ beyond the bar, and a high skewness area (skew($V_\textnormal{los}$) $\geq 0.25$) that spreads from $\ell<0^\circ$ to the nearest bar edge. A third remarkable feature is observed close to the Galactic center, where the positive skewness region penetrates the bar approximately 1.0-1.2 kpc away.
\par Similar features can be seen in the Milky Way data. The low-skewness ``banana-shaped'' region is found at closer distances and slightly larger longitudes ($15^\circ\lesssim\ell\lesssim 35^\circ$), with an additional prolongation along the $\ell=30^\circ$ direction. In the approaching side of the Galaxy, we observe the expected positive skewness region, albeit with a different shape. We also find an extension across the bar two times larger than the one predicted by the barred model.

\par The axisymmetric model predicts a quite uniform distribution with skewness values from $-0.75$ to $0$, which contrast with the positive skewness region seen in the barred model and in the observations. Apart from the negatively skewed area (skew($V_\textnormal{los}$) $\leq -0.5$), no other estructures are observed in the axisymmetric model.

\par The last rows in Figs. \ref{Pict_Solid_full_ld} and \ref{Pict_Contourf_full_ld} show the maps of the kurtosis for models and observations. It is clear from the non-axisymmetric model that the minimum kurtosis region ($<-1.0$) corresponds to the bar, while the maximun values ($>1.0$) are found in the low-skewness region. On the contrary, the unbarred simulation shows positive kurtosis everywhere, with an enhancement at $\ell\approx15^\circ$ and $d_b\gtrsim7.0$ kpc probably caused by a spiral arm. It is important to note that both models differ in the general trend of the kurtosis, with positive (negative) values for the axisymmetric (barred) model. This is additional proof of the presence of the Milky-Way bar, since the observational data show negative or almost-zero kurtosis values for $\ell<20^\circ$. Furthermore, the barred model is supported by the observation of the high kurtosis region related to the receding bar edge.
%
%
%
\subsection{Distributions of the line-of-sight velocities}
\par We compared the $V_\textnormal{los}$ distributions of the models to the observations to get a deeper insight on the kinematics. We now select the regions where prominent signatures of the presence of the bar are observed (see Section \ref{subsec_skew_kurt}) and calculate their $V_\textnormal{los}$ distributions. The regions are the same for all the maps to avoid an artificial concordance between the distributions of separated regions, although this method may distort the results due to the mix of sources with different kinematics. 
\par
\begin{figure*}
\begin{center}
\includegraphics[scale = .50]{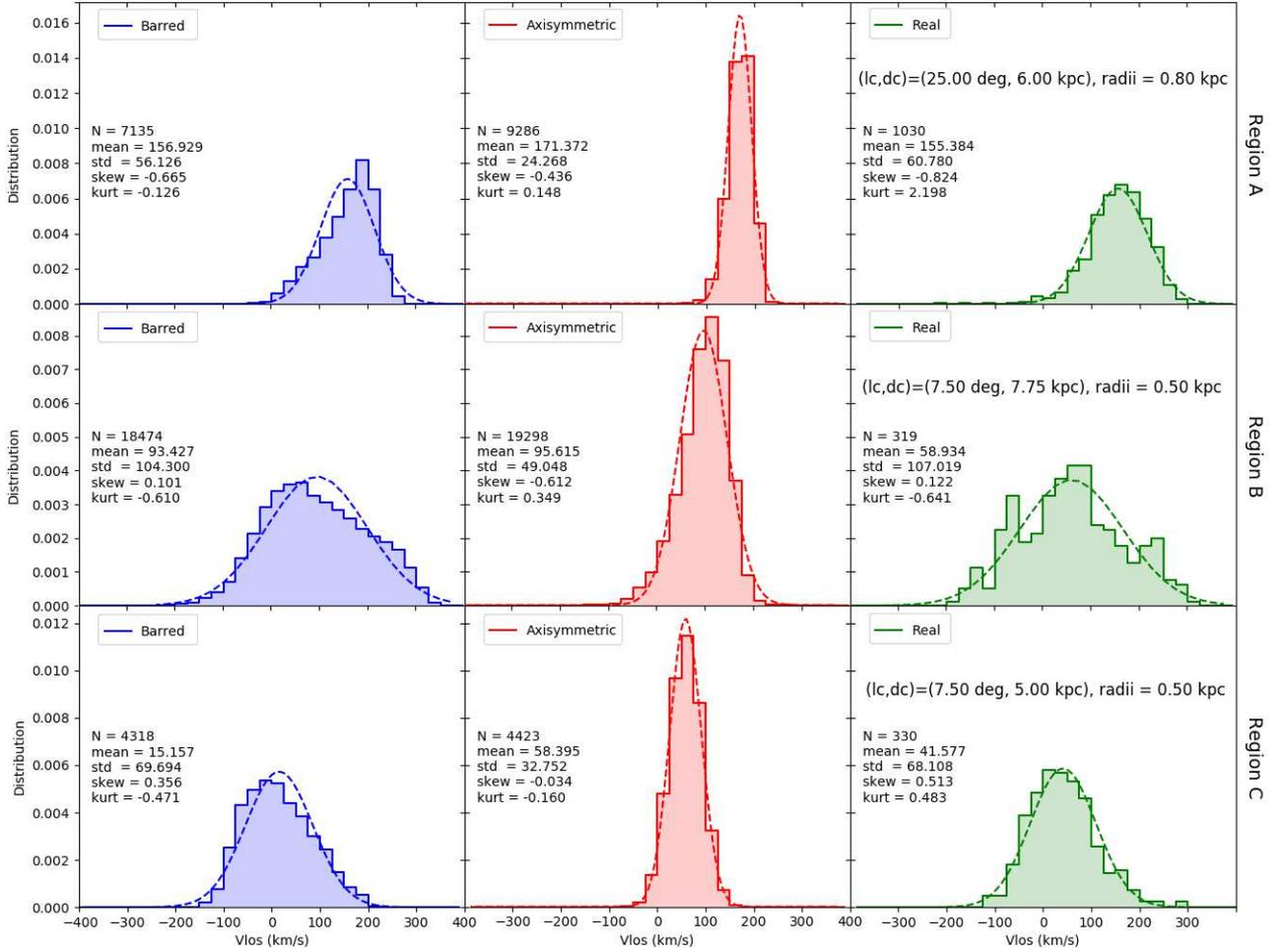}
\caption{Distributions of the line-of-sight velocitites for the three selected regions in the galactic plane ($|Z|<$ 1 kpc). Each distribution spans the range from -400 to 400 km s$^{-1}$ with a bin size of 25 km s$^{-1}$. The dashed lines represent the Gaussian distributions with the same mean value and standard deviation as the binned distributions. The statistic parameters are detailed in the insets as follows: number of stars (particles for the simulations), mean value, standard deviation, skewness and kurtosis. }
\label{Pict_BCD}
\end{center}
\end{figure*}

 The distributions are plotted in Fig. \ref{Pict_BCD}. The region labelled as A corresponds to the zone of low-skewness tagged in Figs. \ref{Pict_Solid_full_ld} and \ref{Pict_Contourf_full_ld}, Region B encloses the observed extension near the Galactic center, and C is related to the high skewness values at $\ell>0^\circ$. The dashed lines in Fig. \ref{Pict_BCD} represent the Gaussian curves with the same mean and dispersion as the distributions. As can be seen, the axisymmetric model predicts almost Gaussian distributions, while the observations and the barred model show significant deviations from gaussianity. In Region A, for example, both observational data and the barred model show a non-symmetric distribution whose peak is displaced towards values larger than the average (the maximum of the Gaussian curve). In Region B, the distributions tend to be wider and only the axisymmetric model is negatively skewed.
\par A sense of the quantitative agreement between the barred model and the observational data can be obtained by noting that in both cases the Gaussian distribution overestimates the number of sources within the range 100-200 km s$^{-1}$ and subestimates the contribution for $V_\textnormal{los}\gtrsim200$ km s$^{-1}$, where the Milky Way data show a minor peak. On the other hand, the main discrepancies are related to the left half of the distributions, since the secondary peak at $\sim-80$ km s$^{-1}$ is not predicted by the simulations. In Region C, the $\textnormal{V}_{los}$ distribution of the axisymmetric model is nearly Gaussian with an skewness and kurtosis excesses close to zero, while the barred galaxy and the observations are positive skewed with a more extended tail at larger velocities. 
\par
\begin{figure}
\begin{center}
\includegraphics[scale = .60]{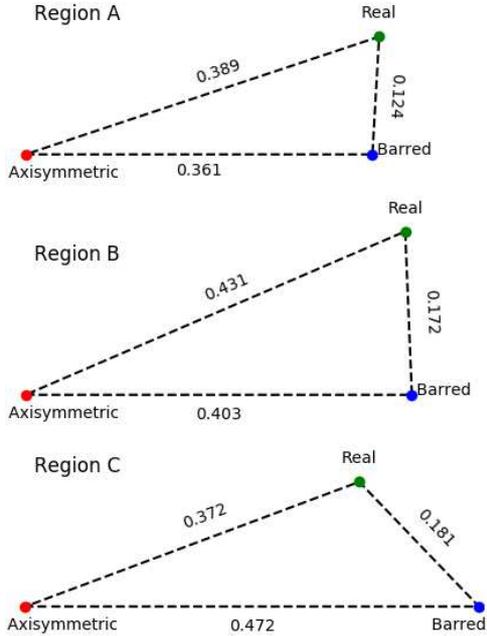}
\caption{Geometric interpretation of the Hellinger distances for the three selected regions (see the text). The horizontal edge accounts for the similarities between the $V_\textnormal{los}$ distributions of both numerical simulations, while the other edges represent the goodness of the description of the observations with the models. All the plots have the same scale. }
\label{Pict_Triangles}
\end{center}
\end{figure}

 In order to quantify the resemblance between the distributions, we have tried different statistical methods such as the Kullback-Leibler divergence \citep{kullback1951} or the Bhattacharyya distance \citep{Bhattacharyya1943}, as well as the widely used $\chi^2$ test. However, we discarded these methods because they do not account for the similarities between the models, and because the Kullback-Leibler divergence and the $\chi^2$ test diverge when the denominator tends to zero. Finally, we have decided to use the Hellinger distance \citep{Hellinger1909} 
 \begin{equation}
 \label{eq_Hellinger}
 H(P,Q) = \sqrt{1-\sum{\sqrt{P_i Q_i}}}
\end{equation}
because it satisfies the triangle inequality, that allows a geometrical representation in which the models correspond to the vertex of a triangle, and the edges are the Hellinger distances between them, which sizes are an indicator of the proximity of the distributions (see Fig. \ref{Pict_Triangles}). This interpretation accounts for the similarities between the models (horizontal lines in Fig. \ref{Pict_Triangles}) that are ignored in other goodness-of-fit estimators such as the Bayes Factor or the Pearson's $\chi^2$ test, and makes it possible to evaluate the quality of the observational data fitting in both absolute (lengths of the edges) and relative terms (edges ratio). For example, in Region A the axisymmetric and the barred models reproduce the observations with similar accuracy (as measured by the Hellinger distance), whereas the barred model shows a Hellinger distance with the Milky Way data three times shorter. A similar result is found for Region B, where the non-axisymmetric model matches better the observed $V_\textnormal{los}$ distribution while all the Hellinger distances have grown, and the axisymmetric model is closer to the barred simulation than to the observations. In Region C, however, this trend is inverted.
\par
\begin{figure}
\begin{center}
\includegraphics[scale = .60]{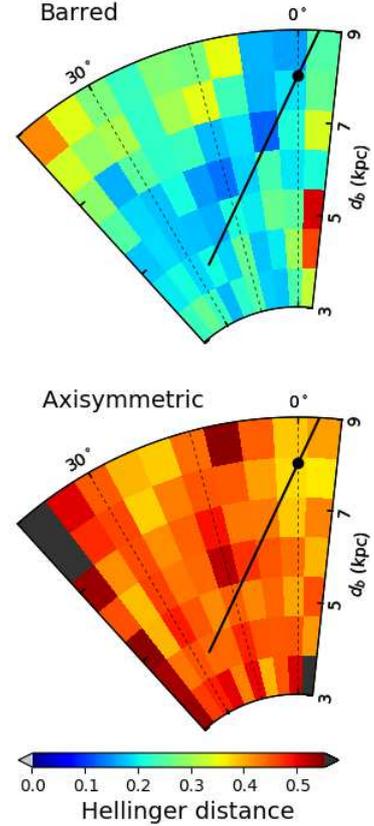}
\caption{Maps of the Hellinger distance for the barred (upper) and the axisymmetric (lower) model. The minimum distances are found along the bar direction only in the non-axisymmetric model, while the unbarred one shows no structure. The Galactic center is represented by the solid black circle at $\ell=0^\circ$, $d_b=R_0=8.0$ kpc, while the solid line represents a 4.5 kpc half-length bar with $\phi=25^\circ$.}
\label{Pict_HellingerMaps}
\end{center}
\end{figure}

 We repeat the analysis over the whole sample and calculate the Hellinger distance in each bin (Fig. \ref{Pict_HellingerMaps}). The $V_\textnormal{los}$ distributions predicted by the barred model tend to be closer to those found in the Milky Way, especially along the bar direction where Hellinger distances are smallest ($\lesssim$0.25). On the contrary, the axisymmetric model shows larger Hellinger distances without any noticeable structure.
\par
\begin{figure}
\begin{center}
\includegraphics[scale = .50]{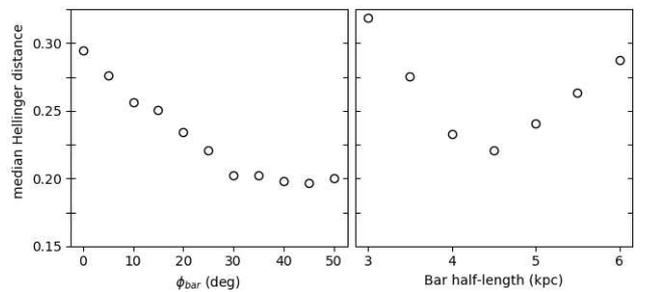}
\caption{Values of the median Hellinger distance as a function of the orientation angle $\phi_{bar}$ (left panel) and as a function of the bar half-length (right panel).}
\label{Pict_BestM}
\end{center}
\end{figure}

 In order to find the best-matching model for the observations, we have studied the impact of the orientation angle $\phi_{bar}$ and the bar half-length on the median Hellinger distance. The orientation angle is varied from $0^\circ$ to $50^\circ$ (step $5^\circ$), while the bar half-length ranges from $3$ to $6$ kpc (step $0.5$ kpc). We observe a decreasing trend in the median Hellinger distance with $\phi_{bar}$, which becomes nearly flat for angles larger than $30^\circ$ (Fig. \ref{Pict_BestM}, left panel). This makes it difficult to constrain the best value of $\phi_{bar}$, although an orientation angle of $\sim30^\circ-35^\circ$ is compatible with some previous estimations (see \citealt{BHGerhard2016} and  references therein). On the other hand, the median Hellinger distance as a function of the bar half-length (Fig. \ref{Pict_BestM}, right panel) shows a minimum at $4.5$ kpc (i. e. the value initially assumed).
%
%
%
%
%
%
%
%
%
%
%
%
%
%
%
%
%
\begin{figure*}
\begin{center}
\includegraphics[scale = .62]{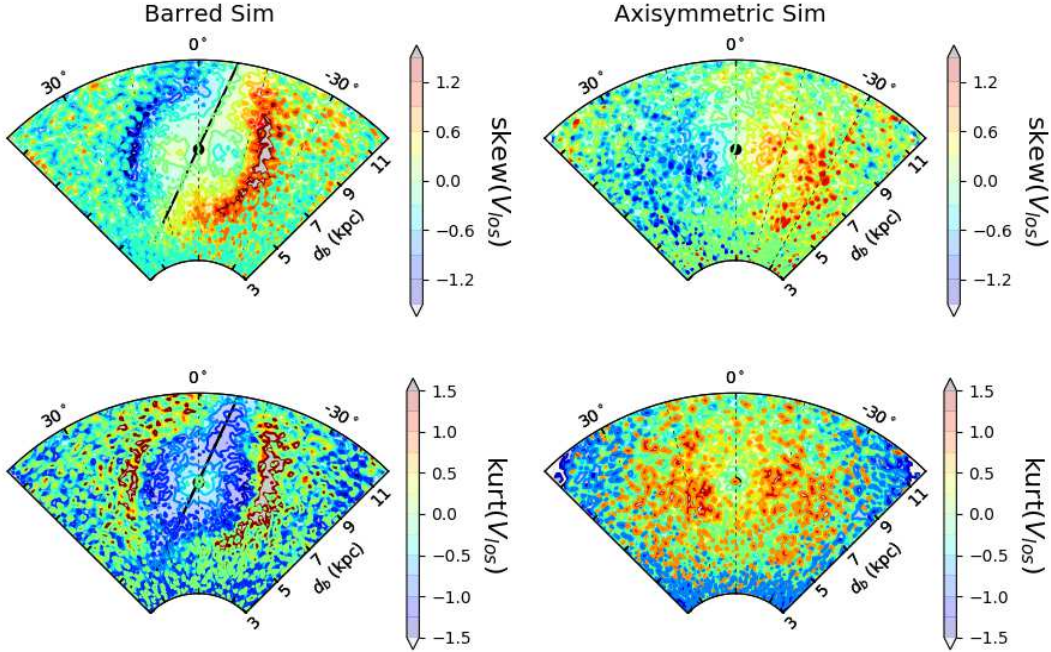}
\caption{Maps for the skewness (first row) and kurtosis (second row) in the ideal case. As in the previous figures, the left (right) column corresponds to the barred (axisymmetric) model. The Galactic center is placed at $\ell=0^\circ$, $d_b=R_0=8.0$ kpc (solid black circle) and the bar is plotted with an orientation angle of $25^\circ$ (black line).}
\label{Pict_Ideal_ld}
\end{center}
\end{figure*}
\section{Discussion}
\label{Discussion_Section}
\par We find signatures of the Galactic bar using the line-of-sight velocities $V_\textnormal{los}$ provided by the new APOGEE DR14 dataset \citep{preprintAPOGEEDR14} and the distance estimations of \citet{BPG_distances, Queiroz_et_al17}.
\par Numerical models of barred galaxies predict a sharp transition in $\langle V_\textnormal{los}\rangle$ at the bar edges \citep{MyPaper} which is not observed in Figs. \ref{Pict_Solid_full_ld} and \ref{Pict_Contourf_full_ld} for three main reasons. Firstly, this transition is clearer if the solar motion is included, which is not our case. Secondly, large bins are required to obtain accurate statistics with this limited dataset. As a consequence of binning, different galactic regions are mixed and their features blurred. Finally, an orientation angle of $20^\circ-30^\circ$ maximizes the similarity between the $\langle V_\textnormal{los}\rangle$ maps for the axisymmetric and barred models. In other words, any other value for the orientation angle would lead to more recognizable bar signatures in $\langle V_\textnormal{los}\rangle$. As a result, the maps for $\langle V_\textnormal{los}\rangle$ are almost identical for the observations and the rescaled simulations (see Section \ref{Sim_data_section}), although the models are intrinsically different.
\par Similarly, the map of the Milky Way $\sigma_\textnormal{los}$ shows a simple decreasing trend with galactocentric distance, compatible with both models (second rows in Figs. \ref{Pict_Solid_full_ld} and \ref{Pict_Contourf_full_ld}), although the barred model is dynamically hotter due to the wider variety of bar orbits. As metal-poor stars are brighter and kinematically hotter, an observational bias in favour of those stars may reproduce the high $\sigma_\textnormal{los}$ seen in the distant regions. However, we discard this selection effect as cause of such structures because the $\langle V_\textnormal{los}\rangle$ maps are similar for all the considered metallicity ranges (Fig. \ref{Pict_AllFeH}), and because the lack of metal-rich stars is only significant beyond the Galactic center.
\par As Figs. \ref{Pict_Solid_full_ld} and \ref{Pict_Contourf_full_ld} show, we can discern three main structures in the skewness distribution with a single imprint on the kurtosis map. These features are nothing but the degradation expected due to our coarse binning (Fig. \ref{Pict_Ideal_ld}).
\par The first structure is a low-skewness and high-kurtosis region within $15^\circ\lesssim \ell \lesssim40^\circ$ and $d_b\gtrsim3$ kpc, whose counterpart in the barred model is found at farther distances $d_b\gtrsim5$ kpc, probably due to an incorrect estimation of distances, with no parallel structure in the axisymmetric case. According to the barred model, this feature appears to be associated with the receding part of the bar, and its position depends strongly on the orientation angle $\phi_\textnormal{bar}$, the bar length, and $R_0$.
\par Another signature of the bar is the positive or near to zero skewness region across the major axis ($d_b\approx7-8$ kpc). In contrast to the previous feature, both the data and the barred model show this enhancement at the same position. We verified that the enhancement is independent of the bar orientation. As Fig. \ref{Pict_Ideal_ld} shows, an analogous feature with $\textnormal{skew}(V_\textnormal{los})<0^\circ$ is expected at negative longitudes. Unfortunately, this part of the sky is not covered by the APOGEE survey due to the geographic location of the Apache Point Observatory, but the ongoing APOGEE-2 project \citep{APOGEE2paper, APOGEE2S_target} or the \textit{Gaia} mission, for example, will provide the data required to verify our hypothesis.
\par The high skewness domain ($>0.25$) observed at positive longitudes constitutes a third evidence for the bar, and can be used to constrain the orientation angle. As our tests with the barred model show, this structure corresponds to the approaching side of the bar, and it is bounded by the receding major axis (providing that $\phi_{bar}$ lies between $0^\circ$ and $\sim60^\circ$), so a small $|\phi_{bar}|$ is highly improbable (Fig. \ref{Pict_Contourf_full_ld}). We attribute the differences in $\ell<0^\circ$ to a relatively low number of sources available ($\lesssim$ 100 stars per bin). It is important to emphasize that the positive skewness features are located in a region where the axisymmetric model predicts negative skew values, so a different $\sigma_{los}$ would not wipe out these structures. 
\par Apart from the previous features, we find good agreement between the kurtosis map of the barred simulations and the observations. As illustrated in the last row of Fig. \ref{Pict_Contourf_full_ld}, the barred model reproduces approximately the regions of opposite sign observed beyond the bar major axis: at $0^\circ \lesssim \ell \lesssim15^\circ$ we find a low kurtosis region (kurt($V_{los}$)$\lesssim-0.5$) while at $15^\circ \lesssim \ell \lesssim  30^\circ $ the kurtosis is clearly positive (kurt($V_{los}$)$>0.5$). These regions contrast with the more homogeneous map predicted by the axisymmetric model, in which the kurtosis is predominantly positive. At closer distances, on the contrary, the observations show no clear pattern.
\par It is worthwhile that all of the structures described above are robust against changes in the distance estimates. Indeed, assuming simply a common age of 5 Gyrs for all the APOGEE DR14 stars to interpolate in the \textsc{parsec} isochrones\footnote{\href{url}{http://stev.oapd.inaf.it/cgi-bin/cmd} } \citep{Bressan_et_al2012}, and adopting the extinction corrections given by \citet{Zasowski2013}, we can estimate the distances of the non-APOGEE DR13 stars \citep{APOGEEDR13paper}. As a result, even though we observe minor discrepacies in the structures compared to those in Figs. \ref{Pict_Solid_full_ld} and \ref{Pict_Contourf_full_ld}, the general pattern does not change substantially, and no new features appear.
\par Our maps are in good agreement with those obtained by \citet{Zasowski_et_al16}, who explored the high-order moments of $V_\textnormal{los}$ for more than 19,000 APOGEE DR12 stars to constrain models of the Milky Way. Although they use a sky projection (i. e. the $\ell$-$b$ plane) to display their results, we find important concordances with our $\ell$-$d_b$ maps, such as the previously described high and low skewness areas, associated with Regions C and A, respectively. As can be seen in their figure 6, they find a negative skewness region at $\ell\approx30^\circ-35^\circ$ which is fully compatible with our Region A, and a peak in the positive skewness area at $\ell\approx10^\circ$ as in Figs. \ref{Pict_Solid_full_ld} and \ref{Pict_Contourf_full_ld}. On the other hand, a major discrepancy appears in the fourth order moment, since they report an almost longitudinally flat distribution for kurt($V_\textnormal{los}$) while we observe a clear enhancement along the $\ell\approx30^\circ$ direction. This noteworthy difference is probably due to the ``cone effect'' characteristic of the $\ell$-$b$ projection, since it mixes stars from different Galactic regions whose kinematics also differs (see their figure 16). We can exclude small number statistics as a source of error in our estimation of kurt($V_\textnormal{los}$) given that this region is one of the most populated areas.
\par In a similar work, \citet{Zhou_et_al2017} fit the line-of-sight velocity distributions with the Gauss-Hermite series proposed by \citet{van_der_Marel1993}:
\begin{equation}
 \label{GaussHermite}
 \frac{\gamma e^{-(V-\bar{V})^2/{2\sigma^2}} }{\sqrt{2\pi}\sigma}\left[ 1+h_3 H_3\left(\frac{V-\bar{V}}{\sigma}\right)+h_4 H_4\left(\frac{V-\bar{V}}{\sigma}\right)\right]
\end{equation}

where $H_3$ and $H_4$ are the third and fourth order Hermite polynomials, respectively, and $\gamma$, $\bar{V}$, $\sigma$, $h_3$ and $h_4$ are the free parameters of the fit. In particular, the coefficient $h_3$ ($h_4$) accounts for the asymmetric (symmetric) deviation from a Gaussian distribution. Their results show a positive $h_3$-$\bar{V}$ correlation in the bar region ($|\ell|<10^\circ$), in contrast to the anticorrelation found in disc-dominated areas ($|\ell|>10^\circ$), supporting the predictions of \citet{Bureau_Athana_2005}, \citet{Shen_Debattista2009}, or \citet{Iannuzzi_Athana_2015}. In our analysis of $h_3$, however, we do not find similar trends with $\bar{V}$ for two main reasons. First, the face-on projection considered in our maps automatically excludes the contribution of foreground disc stars to the bar region. This is opposite to the case for the edge-on projection, in which foreground disc stars play an important role in the statistics for the central longitudes. For example, \citet{Zhou_et_al2017} report a weaker $h_3$-$\bar{V}$ correlation in the bar region when the foreground disc stars are excluded in the model of \citet{Shenetal2010}, and a shift in the maximum of $h_3$ once the $T_{eff}<4000$ K cutoff\footnote{\citet{Zhou_et_al2017} propose a cutoff in $T_{eff}$ due to the lack of distance estimations for some APOGEE sources (see their figure 1.b). } is applied to the APOGEE data (see their figure 4). Second, different projection choices tend to group sources in different ways, leading to distributions with different moments and best-fitting parameters, affecting the relations among them; such as the $h_3$-$\bar{V}$ correlation, which may appear distorted or even removed. This can be seen in the recent work of \citet{Li_et_al2018}, in which the inclination of the models dramatically changes the $h_3$-$\bar{V}$ relationship, especially in the outer parts of the bar, where the correlation can be inverted. Finally, the $h_3$-$\bar{V}$ correlation reported in the edge-on view is supported by the gradual increase in $\bar{V}$ with longitude, which al low $|b|$ can be studied it as a simple $h_3(\ell)$ relation. This is not the case for the face-on view, since $h_3$ and $\bar{V}$ also vary with distance.
\par The comparison of the $V_\textnormal{los}$ distributions (Fig. \ref{Pict_BCD}) supports our interpretation of the maps. In quantitative terms, the barred model tends to predict the mean value, the dispersion, and the skewness of $V_\textnormal{los}$ better than the unbarred model, even though the shape of the distributions differs as discussed for the case of Region A. On the other hand, no model reproduces the observed kurtosis in Regions A and C, but in Region B the barred simulation predicts a kurtosis of -0.610, while the observed value is -0.641.
\par In Region B the distributions tend to be wider due to the high velocity dispersion of the Galactic center. The barred simulation and the observational data show a non-Gaussian distribution whose complexity cannot be explained by an axisymmetric galaxy, with a common over and underestimation of the sources with respect to the Normal distribution (dashed lines). The observational data show two secondary peaks at $V_\textnormal{los}\approx225$ km s$^{-1}$ and at $V_\textnormal{los}\approx-75$ km s$^{-1}$. The former is analogous to that reported by \citet{NideveretalHVP2012} in APOGEE commissioning data, and later confirmed by \citet{Zasowski_et_al16} and \citet{Zhou_et_al2017} using APOGEE DR12 and DR13 data, respectively. In particular, the high velocity peaks observed at $(\ell,b)=(6^\circ,0^\circ), (10^\circ, \pm 2^\circ)$ are statistically significant \citep{Zhou_et_al2017}. We find that the local maximum at $V_\textnormal{los}\approx225$ km s$^{-1}$ is compatible with these previous works for several reasons:
\begin{enumerate}
 \item It is observed in a histogram with similar bin size ($25$ km s$^{-1}$) to that used by \citet{NideveretalHVP2012} and \citet{Zhou_et_al2017} ($20$ km s$^{-1}$). This is important, because a larger bin size reduces resolution and distorts the shape of the distribution, even though the Poissonian noise is reduced.
 \item Since Region B spans the longitudinal range from $3.8^\circ$ to $11.2^\circ$, it contains six of the eight bulge fields where the high velocity peaks were observed, including those at $(\ell,b)=(6^\circ,0^\circ), (10^\circ, \pm 2^\circ)$. Moreover, Region B is centred at $\sim1.1$ kpc from the Galactic Center, which is consistent with the kiloparsec-scale nuclear disc proposed by \citet{DebattistaNessEarpHVP2015}.
 \item According to \citet{NideveretalHVP2012} and \citet{Zhou_et_al2017}, the stars in the high velocity peaks do not show chemical differences with respect to the main peak. We have verified that there is no relationship between $V_{los}$ and [Fe/H] in the stars of the Region B.
\end{enumerate}

\par Although \citet{Zhou_et_al2017} report negative velocity peaks similar to that we found, they conclude that most of them are due to statistical fluctuations within $\sim2\sigma$ the Poissonian error. In our case, this negative velocity peak may be a consequence of the non-uniform sky coverage of the APOGEE survey, since its stars lie in the half of Region B closer to the Galactic center. However, according to the statistical criteria adopted in \citet{Zhou_et_al2017}, it can be also considered a result of the Poissonian noise since it is found at $\sim 2.02 \sigma$.
\par The $V_\textnormal{los}$ distributions in Region C seem to be more normally distributed, but only the barred model and the observational data share certain non-Gaussian features, such as the long tail at large $V_\textnormal{los}$ and the overestimation in the number of sources between 0 and -75 km s$^{-1}$. In Region A, the barred simulation provides the best fitting to the $V_\textnormal{los}$ distribution of the Milky Way (upper panel in Fig. \ref{Pict_BCD}) despite the location of their low-skewness regions differs. Furthermore, our tests with different barred models prove that the shape of the distribution in this region changes with the orientation angle of the bar, the Sun-Galactic center distance $R_0$, the pattern speed, and the length of the bar.
\par According to the Hellinger distances, the barred model provides the best description for the Milky Way kinematics in the three regions. As we can see in Fig. \ref{Pict_Triangles}, the distances between the axisymmetric model and the observation are at least two times larger than those between the observations and the barred model. We must recall that with this Hellinger distance test we are not trying to account for the differences in the position of the structures, but just compare the distributions of certain regions.
%
%
%
\section{Conclusions}
\label{Summary_Section}
\par Both the barred and the axisymmetric models can reproduce the maps of $\langle V_\textnormal{los}\rangle$ and the trend in $\sigma_\textnormal{los}$ observed in the Milky Way. Our results show, however, that it is possible to discern structures in the maps of higher-order moments of $V_\textnormal{los}$ that can not be explained with an unbarred galaxy. We identify a structure in the Milky Way data ubiquitous to barred galaxies (Region B), as well as two other features that depend on the orientation angle $\phi_\textnormal{bar}$ (Regions A and C). Unlike previous studies, we detect an enhancement in kurtosis associated with the end of the bar that breaks the uniformity of the map. 
\par We compare the distributions of $V_\textnormal{los}$ in three regions and find good agreement between the barred model and the observational data. In order to quantify the similarity, we introduce the Hellinger distance to measure the proximity between two distributions, which supports our interpretation of the maps. In addition to that, the study of the $V_\textnormal{los}$ distribution in Region B reveals a secondary peak similar to those reported by \citet{NideveretalHVP2012}, and another local maximum which may be caused by limitations in sky coverage of APOGEE, and will be solved with APOGEE-2S observations \citep{APOGEE2S_target}.
\par The APOGEE-2 project or the \textit{Gaia} mission will provide the line-of-sight velocities needed to confirm our interpretation of the central structures. Furthermore, we plan to extend the quantitative comparison using the Hellinger distance to barred models with different evolutionary paths, and of different bar shapes, lengths, and pattern speeds.
\section*{Acknowledgments}
C.A.P. is thankful to the Spanish Ministry of Economy and Competitiveness (MINECO) for support through grant AYA2014-56359-P. CDV and PAP acknowledge financial support from MINECO through grant AYA2014-58308-P. CDV also acknowledges financial support from MINECO through grant RYC-2015-18078. DAGH and OZ acknowledge support provided by the Spanish MINECO under grants AYA-2014-58082-P and AYA-2017-88254-P. 

We acknowledge the contribution of Teide High-Performance Computing facilities to the results of this research. TeideHPC facilities are provided by the Instituto Tecnológico y de Energías Renovables (ITER, SA). URL: \url{http://teidehpc.iter.es}.

Funding for the Sloan Digital Sky Survey IV has been provided by the
Alfred P. Sloan Foundation, the U.S. Department of Energy Office of
Science, and the Participating Institutions. SDSS acknowledges
support and resources from the Center for High-Performance Computing at
the University of Utah. The SDSS web site is \url{www.sdss.org}.

SDSS is managed by the Astrophysical Research Consortium for the Participating Institutions of the SDSS Collaboration including the Brazilian Participation Group, the Carnegie Institution for Science, Carnegie Mellon University, the Chilean Participation Group, the French Participation Group, Harvard-Smithsonian Center for Astrophysics, Instituto de Astrofísica de Canarias, The Johns Hopkins University, Kavli Institute for the Physics and Mathematics of the Universe (IPMU) / University of Tokyo, Lawrence Berkeley National Laboratory, Leibniz Institut für Astrophysik Potsdam (AIP), Max-Planck-Institut für Astronomie (MPIA Heidelberg), Max-Planck-Institut für Astrophysik (MPA Garching), Max-Planck-Institut für Extraterrestrische Physik (MPE), National Astronomical Observatories of China, New Mexico State University, New York University, University of Notre Dame, Observatório Nacional / MCTI, The Ohio State University, Pennsylvania State University, Shanghai Astronomical Observatory, United Kingdom Participation Group, Universidad Nacional Autónoma de México, University of Arizona, University of Colorado Boulder, University of Oxford, University of Portsmouth, University of Utah, University of Virginia, University of Washington, University of Wisconsin, Vanderbilt University, and Yale University.
%
%

\bibliographystyle{mnras}
\bibliography{Stellar_kinematics.bib}
\end{document}